\documentclass[runningheads]{svmult}

\usepackage{makeidx}   % allows index generation
\usepackage{graphicx}  % standard LaTeX graphics tool
\usepackage{subeqnar}  % subnumbers individual equations
\usepackage{multicol}  % used for the two-column index
\usepackage{physprbb}  % modified textarea for proceedings,
\makeindex             % used for the subject index
\begin{document}
\title*{VLT-ISAAC 3--5\,$\mu$m spectroscopy of low-mass young stellar
objects: prospects for CRIRES}
\titlerunning{ISAAC 3--5\,$\mu$m spectroscopy of YSOs}
% allows abbreviation of title, if the full title is too long
% to fit in the running head
%
\author{Klaus M. Pontoppidan\inst{1}
\and Ewine F. van Dishoeck\inst{1}}
\authorrunning{Pontoppidan \& van Dishoeck}
% if there are more than two authors,
% please abbreviate author list for running head
%
%
\institute{Leiden Observatory, P.O. Box 9513, NL-2300 RA Leiden, Netherlands}

\maketitle              % typesets the title of the contribution

\begin{abstract}
We present results from an extensive spectroscopic survey in the 3-5\,$\mu$m wavelength region of low-mass young stellar objects using VLT-ISAAC. Medium resolution spectra ($\lambda/\Delta\lambda\sim 1\,000-10\,000$) of young embedded stars in the mid-infrared allow for detailed studies of
ro-vibrational lines from molecular gas, interstellar ices and Polycyclic Aromatic Hydrocarbons (PAHs). By taking advantage of this wide range of molecular tracers available within a few spectral settings, the survey has helped to constrain the chemical evolution of cold molecular material in low-mass star forming regions as well as the physics of disks surrounding protostars. In this contribution, we will review the various spectral diagnostics of molecular material, which require ground-based high resolution infrared spectroscopy. The importance of a high resolution spectroscopic capability as will be offered by CRIRES is discussed in the context of the physics and chemistry of low-mass star formation.

\end{abstract}

\section{Introduction}

The earliest stages of star- and planet formation are characterised by
enormous amounts of gas and dust (up to several hundred mag of
extinction) on scales of a few thousand AU. Since the young star
itself is often invisible, spectroscopic observations of the
surrounding material are the only way to probe the physical and
chemical processes that take place deep inside the circumstellar
envelope. High resolution infrared spectroscopy is a particularly
powerful tool to study these early phases, and gives information that
is quite complementary to that obtained from submillimeter emission
lines.

The atmospheric windows from 3--5\,$\mu$m contain a range of important
spectroscopic tracers of circumstellar material. These include
vibrational bands of solid-state volatiles, such as water ice at
2.8--3.8\,$\mu$m, $^{12}$CO ice at 4.67\,$\mu$m and $^{13}$CO ice at
4.78\,$\mu$m in addition to weaker bands from solid CH$_3$OH, H$_2$CO,
OCN$^-$, OCS. Also the C-H stretching band from Polycyclic Aromatic
Hydrocarbons (PAHs) at 3.3\,$\mu$m is easily observable. A vast number
of ro-vibrational lines from gas-phase molecules are in principle
present throughout the mid-infrared wavelength range. The bands most
easily accessible from the ground are the CO overtone bandheads around
2.3\,$\mu$m and the CO fundamental band around 4.67\,$\mu$m, but
also water vapour, CH$_4$ and possibly CH$_3$OH gas can be observed from the
ground, albeit with considerable difficulty \cite{Carr04}\cite{Lacy91}.

While ground-based mid-infrared spectroscopic observations are limited
in sensitivity and wavelength coverage by atmospheric emission and
absorption, respectively, they outperform current space-based
spectroscopy by orders of magnitudes in terms of spatial and spectral
resolution. For example, the recently launched Spitzer Space Telescope
can produce good quality spectra with signal-to-noise ratios
sufficient for ice absorption line studies of sources fainter than
10\,mJy in the 5-20\,$\mu$m wavelength region. However, this is with a
spectral resolving power of only $\lambda/\Delta\lambda\sim100-600$
and a spatial resolution of 2-6", depending on wavelength.  For
comparison, VLT-ISAAC reaches an effective signal-to-noise ratio of
$\sim 10-20$ at 4.7\,$\mu$m for a 100\,mJy source, but with a spectral
resolution of $\lambda/\Delta\lambda\sim$10\,000 and a typical spatial
resolution of 0.2--0.4". VISIR offers similar capabilities in the $N$-
and $Q$-bands and CRIRES will add another order of magnitude to the
spectral resolution in the 3--5\,$\mu$m region. As will be discussed
below, such high spectral resolution is essential to reliably measure
the intrinsically very narrow gas-phase lines as well as observe
narrow substructure in solid-state features.

In this contribution, we present results from 5 years of dedicated $L$-
and $M$-band spectroscopic observations with ISAAC of interstellar gas
and dust in low-mass star-forming regions. Finally, we will discuss
observational strategies for the future.

\section{Summary of observations and importance of high spectral
resolution}

A total of 60 lines of sight in the Ophiuchus, Chamaeleon, Corona
Australis, Vela, Orion and Serpens star-forming regions have been
observed in the 2.8--4.2\,$\mu$m and 4.55--4.8\,$\mu$m regions and 15
lines of sight have been observed in one wavelength region only.  Of
these, 80\% show the presence of ice while 70\% show ro-vibrational
lines from CO. Good quality spectra with signal-to-noise ratios in
excess of 15 were obtained of sources as faint as 20 mJy at
3.5\,$\mu$m and 100 mJy at 4.7\,$\mu$m. All of the ice features are
seen in absorption against the hot ($T_{\rm dust}>300$ K) dust in the
immediate surroundings of the young star, whereas the gas-phase lines
are seen both in absorption and emission.

In general, the 2.8--4.2\,$\mu$m region was observed
using the ISAAC low resolution module giving resolving powers of
$R=\lambda/\Delta\lambda=$600--1200, depending on the slit width. This
is sufficient for getting high quality spectra of the broad 3\,$\mu$m
water ice band. However, there are instances where higher spectral
resolution significantly improves the quality of solid state spectra,
especially in wavelength regions dominated by complex telluric
absorption.  Telluric absorption lines are intrinsically narrow ($\sim
5\,\rm km\,s^{-1}$) and therefore always unresolved with the present
instrumentation. By going to higher spectral resolution, it becomes
easier to identify the sections of the spectrum where spectral
information is lost or heavily affected due to strong or saturated
telluric lines. For the interpretation of shallow substructure in ice
bands, this becomes increasingly relevant.  For instance, in order to
properly confirm detections of the weak 3.53\,$\mu$m band from solid
methanol (CH$_3$OH), a spectrum using the high resolution module was
often taken to improve the telluric correction at
$R$=3000--6000. Another example is the 3.3\,$\mu$m PAH feature which
is often difficult to separate from the 3.30\,$\mu$m Pf $\delta$
hydrogen recombination line in the low resolution module.

\begin{figure}[hb]
\begin{center}
\includegraphics[height=.9\textwidth, angle=90]{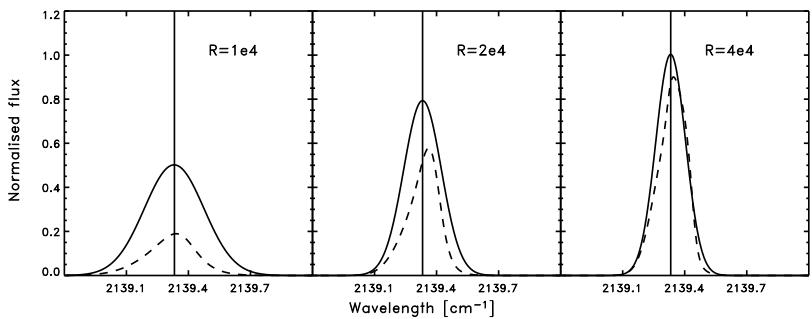}
\end{center}
\caption[]{A simulation of the effect on the recovered profile of an
unresolved line by simple division with a telluric standard star. The
full line shows the intrinsic line profile as it should appear when
observed at the relevant resolution. The dashed line shows the profile
obtained after standard correction for the telluric absorption
spectrum. The specific case shown is for the $^{12}$CO P(1) v=1-0 line
at 2139.43\,cm$^{-1}$ for a line width of 15\,$\rm km\,s^{-1}$ and a
Doppler shift of 10\,$\rm km\,s^{-1}$. A telluric spectrum observed at
a resolving power of $\lambda/\Delta\lambda=100\,000$ was used for the
calculation.}
\label{TellExample}
\end{figure}

In the $M$-band at 4.5--5.0\,$\mu$m, we found that the highest
resolution available with ISAAC of R=10\,000 gave the best
result. This was partly due to the presence of many ro-vibrational
lines from gaseous CO and partly due to the presence of narrow
substructure in the 4.67\,$\mu$m CO ice band. The circumstellar CO gas
phase lines are known in some cases to be as narrow as a few $\rm
km\,s^{-1}$, i.e. much less than the best resolution of ISAAC of
30\,$\rm km\,s^{-1}$.  Traditionally, telluric lines are removed by
dividing with a spectrum of an intrinsically featureless standard
star. However, this procedure assumes that the convolution with an
instrument profile of a product of two spectra is the product of each
spectrum convolved with the instrument profile. This is clearly only
valid when the instrument profile is negligible, i.e. when the
features in the spectrum are completely resolved or when the telluric
spectrum is approximately constant.

There are two realistic solutions to this fundamental problem. The
first is to go to higher resolving power which will be offered by
CRIRES, and the second is to observe the spectrum at a time of the
year when the line of interest is shifted away from the telluric
features. The first option is the better, since even reasonably small
variations in the telluric spectrum can shift and distort astronomical
lines.  Fig. \ref{TellExample} shows an example of the effect of
telluric lines on an unresolved CO ro-vibrational line. For a Doppler
shift of 10\,$\rm km\,s^{-1}$, the derived line strength can be more
than a factor of 2 wrong at the best resolution of ISAAC, while a
Doppler shift of 20\,$\rm km\,s^{-1}$ causes smaller, but
non-negligible errors. Clearly, for this type of line, a spectral
resolving power of more than 50\,000 is required to obtain robust line
strengths, centers and profiles.  For our ISAAC sample, the problem is
most clearly illustrated by the almost total absence of any CO
ro-vibrational lines toward sources in the Chamaeleon star-forming
cloud, where the shift at a declination of -77$^\circ$ is always less
than 10 km s$^{-1}$ for nearby molecular clouds. Thus, this should not be interpreted as a real
lack of lines. To ensure that the effect of telluric lines is small,
the velocity shift should be at least one resolution element, or the
astronomical lines should be resolved.

\section{General results}

\subsection{Overview}

An overview of our results, together with illustrative spectra, is
presented in \cite{Messenger03}.  Some of the highlights include the
first detection of solid methanol in low-mass protostars, a key
ingredient for building more complex organic molecules
\cite{Pontoppidan03b}; direct evidence for significant freeze-out in
edge-on circumstellar disks \cite{Thi02}; sensitive limits on minor
ice components such as ammonia and deuterated water
\cite{Dartois02,Taban03,Dartois03}; and a new weak feature at 2175
cm$^{-1}$ (4.61 $\mu$m) which may be due to CO directly bound to the
silicate surface \cite{Fraser04}.

The unprecented combination of high spectral resolution and high $S/N$
has also allowed the solid CO band at 4.67\,$\mu$m to be fully resolved and to be
studied for a large number of sources \cite{Pontoppidan03a}. Surprisingly, excellent fits
to {\it all} our spectra can be obtained using a phenomenological
decomposition into just three components, with only the relative
strength of these components changing from object to object. This
leads to the important conclusion that the CO ice has the same
fundamental structure along all lines of sight.  For most sources, a
significant fraction of the CO is in nearly pure form, i.e., not mixed
with H$_2$O ice.  This presents a puzzle to theories of ice mantle
formation: either the segregation of the CO and other species has
occurred prior or during freeze-out, or subsequent processing of the
ice and selective desorption and recondensation have resulted in
separation of the components.  

\begin{figure}[hb]
\begin{center}
\includegraphics[width=.8\textwidth]{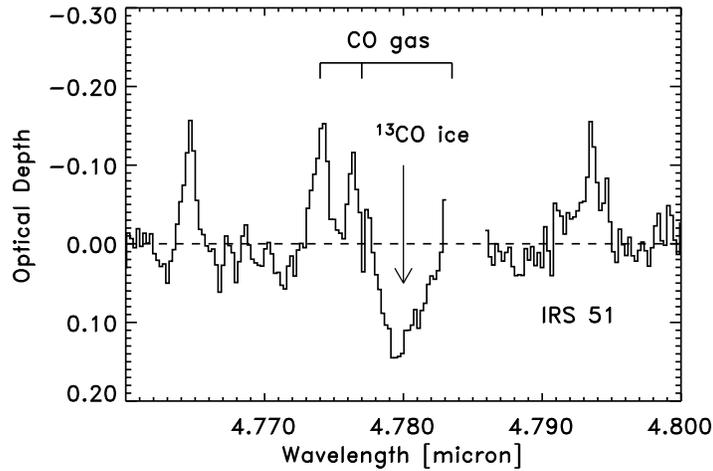}
\end{center}
\caption[]{The $^{13}$CO ice band observed toward IRS 51 in Ophiuchus, from \cite{Pontoppidan03a}.}
\label{13CO}
\end{figure}

In some sources, the $^{12}$CO ice
feature is so strong that its isotopic $^{13}$CO counterpart has been
detected.  Since $^{13}$CO is only a minor component of the ice, its
line shape does not depend on the grain shape and allows further
constraints to be placed on the CO ice environment.
Because the $^{13}$CO ice band is not broadened by grain shape effects, it
becomes extremely narrow and requires resolving powers in excess of $R=10\,000$
in order to be fully resolved. In addition, the band is often blended with ro-vibrational CO gas-phase lines, and a high spectral resolution is essential to disentangle the two phases. 
In Fig. \ref{13CO} is shown an example of a $^{13}$CO ice band.

\subsection{Peculiar sources}
While most 3--5\,$\mu$m spectra of low-mass young stellar objects have
similar characteristics, a few peculiar objects were identified. Here,
two examples are described.

\begin{figure}[hb]
\begin{center}
\includegraphics[width=.8\textwidth]{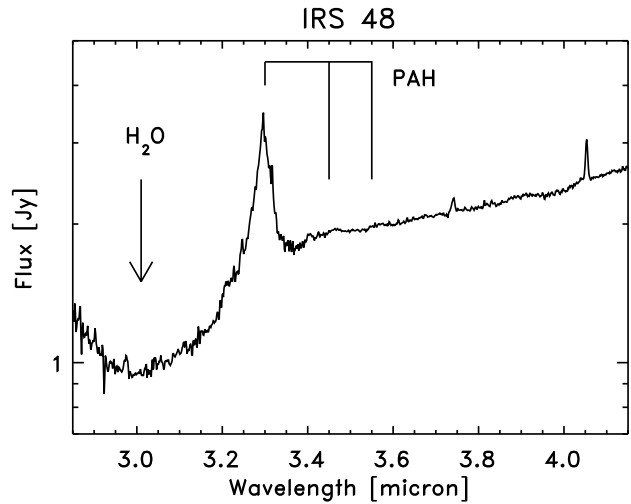}
\end{center}
\caption[]{$L$-band spectrum of IRS 48 showing bright PAH 3.3\,$\mu$m emission and a shallow 3.01\,$\mu$m water ice band in absorption.}
\label{IRS48}
\end{figure}

\paragraph{IRS 48 (WLY 2-48)} Almost none of the observed low-mass YSOs show
compact emission in the 3.3\,$\mu$m band of polycyclic aromatic
hydrocarbons (PAHs). While extended emission from PAHs is seen throughout
molecular clouds and compact emission is seen from disks around some Herbig Ae stars, 
clear PAH features are rarely seen toward low-mass YSOs. 
However, IRS 48 ($L=7.4\,L_{\odot}$) in the $\rho$ Oph cloud has a very strong and compact ($<$80\,AU) 3.3\,$\mu$m PAH band with a peak flux over the continuum of almost 1 Jy. The $L$-band spectrum of IRS 48 is shown in Fig. \ref{IRS48}. IRS 48 is classified as a low-mass flat spectrum YSO \cite{Bontemps01}. It is an interesting question whether the compact PAH emission from embedded protostars is due to a circumstellar disk as that of the more evolved Herbig Ae stars \cite{Boekel04}\cite{Geers04}.

\paragraph{GSS 30 IRS 1} 
This embedded star in the Ophiuchus cloud core exhibits the highest ro-vibrational CO line-to-continuum ratio of our sample. The lines
are spatially extended to a distance of 2"=320\,AU from the central source. These two properties make this source unique among the observed CO emission line sources. Radiative transfer modeling
has shown that the lines can be reproduced by a single temperature gas of a little over 500\,K.
However, to simultaneously model the optical depth, the high line flux level and the observed spatial extent of the line emission, it
was found that lines emitted from the central 20\,AU of a circumstellar disk must be scattered in the bipolar cavity also seen in near-infrared imaging. 
The single temperature can then be explained as a cooling plateau behind a shock, possibly an accretion shock, on the surface of an embedded circumstellar disk. The peculiar geometry of the source thus enabled a study of the physics of a region normally obscured from view by a dense and dusty envelope. However, the lines were not spectrally resolved by ISAAC and only barely resolved at R=25\,000 in subsequent Keck spectroscopy. Thus, this is clearly a case where ground-based high resolution spectroscopy is required to learn more.

\section{CO ro-vibrational lines from embedded young low-mass stars}

The CO ro-vibrational lines observed toward embedded stars show very
diverse structures, ranging from very deep absorption lines to bright
emission lines. Also, the line widths vary from unresolved at
$\lambda/\Delta\lambda =10\,000$ or 20\,$\rm km\,s^{-1}$ to being as
wide as 100\,$\rm km\,s^{-1}$. Some sources show different velocity
components with widely different optical depths and excitation
temperatures in absorption along the line of sight. Disentangling all
the gas components that contribute to the CO ro-vibrational spectra in
many sources clearly requires higher spectral resolution spectroscopy
with CRIRES.

\begin{figure}[b]
\begin{center}
\includegraphics[width=.9\textwidth]{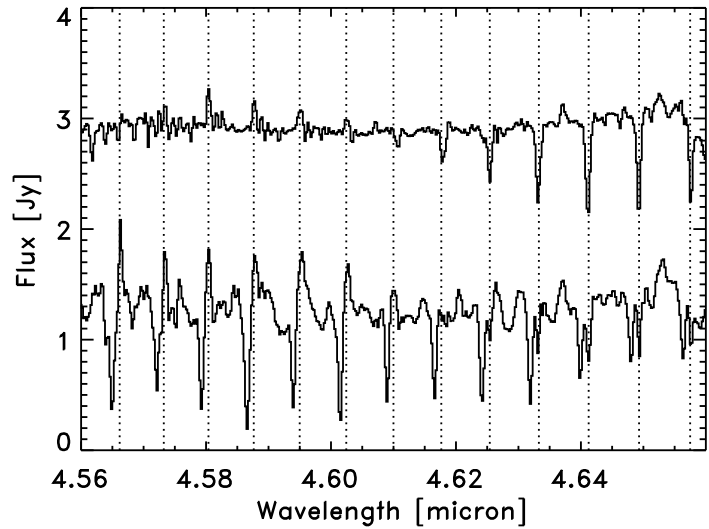}
\end{center}
\caption[]{Comparison of the CO ro-vibrational lines from each
component of the 1.6" binary YSO EC 90 in the Serpens core. The dotted lines indicate
the positions of the ro-vibrational transitions of $^{12}$CO to the
first vibrationally excited level.}
\label{EC90}
\end{figure}

An interesting example of a complex ro-vibrational CO case is
presented in Figure \ref{EC90}, which shows separate 4.6\,$\mu$m
spectra of each component of the low-mass binary YSO EC 90 in the
Serpens core. The two components are separated by only $1.6"=400\,$AU,
yet the line spectra differ dramatically. Both lines of sight show
similar contributions from cool gas near zero velocity while EC 90B
shows an additional warm, narrow component shifted by 80\,$\rm
km\,s^{-1}$.  Finally, broad emission lines are seen from EC 90B near
zero velocity. It is tempting to conclude that the cool component seen
toward both components is due to a circumbinary envelope, while the
complex line structure toward EC 90B is due to activity in the inner
disk of this star, possibly related to an outflow along the line of
sight. It is not clear, however, how a compact clump of gas is
accelerated to a narrow range of velocities around 80\,$\rm
km\,s^{-1}$ as evidenced by the strongly shifted, yet unresolved
lines. Such extremely high velocity clumps or `bullets' have been detected
previously only at millimeter wavelengths in even more deeply embedded
Class 0 objects, e.g.\ \cite{Bachiller91}.

\begin{figure}[ht]
\begin{center}
\includegraphics[width=.9\textwidth]{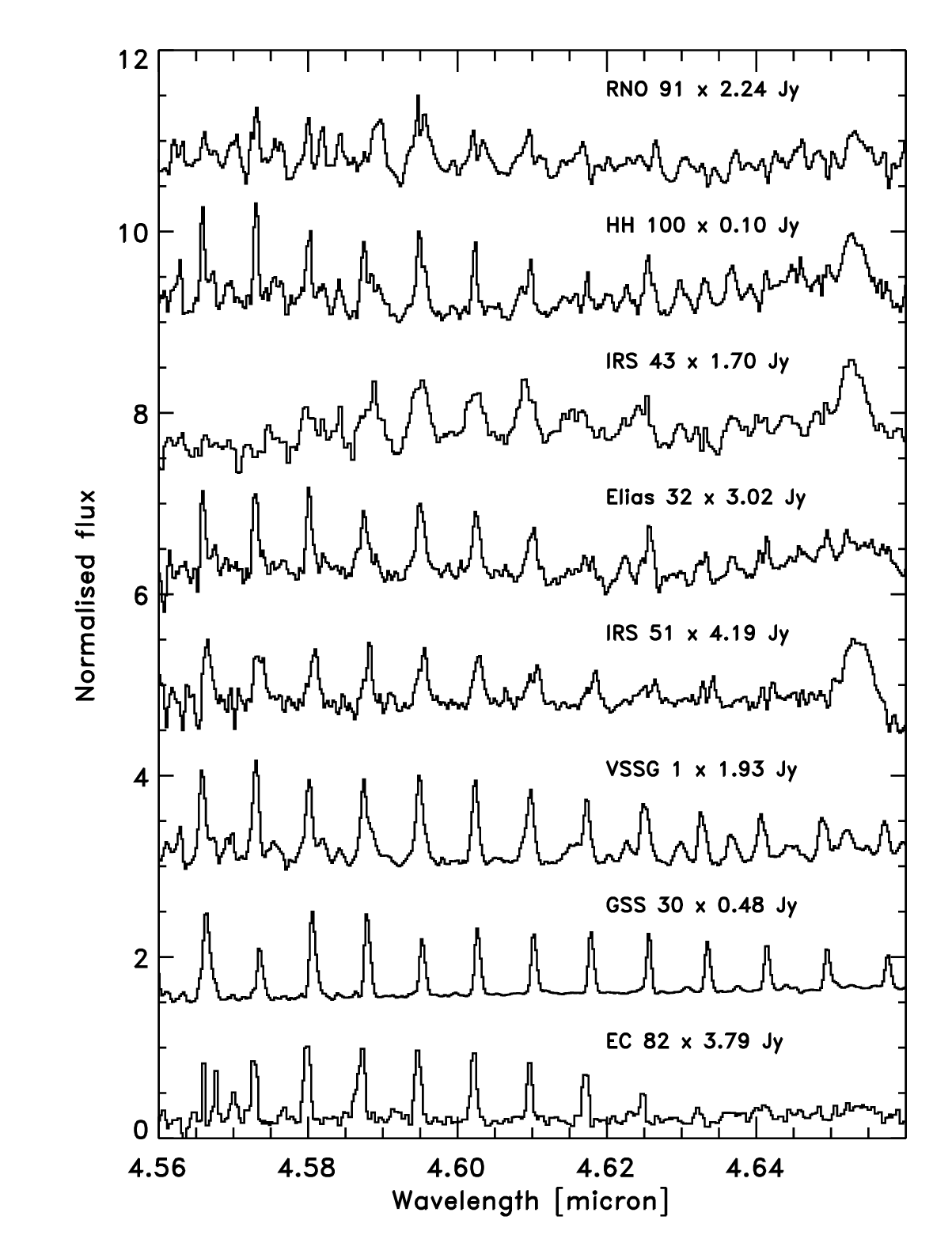}
\end{center}
\caption[]{Gallery of CO ro-vibrational emission lines observed toward
embedded low-mass YSOs ordered according to complexity. The spectra
have been scaled by the amount indicated in the figure and shifted for
clarity. Broad absorption bands from ices have been removed from the
spectra of RNO 91, HH 100, IRS 43, Elias 32 and IRS 51.}
\label{gal}
\end{figure}

Figure \ref{gal} shows a gallery of ro-vibrational emission lines
toward embedded objects. The gallery is ordered according to
increasing complexity of the line spectrum, such that spectra showing
broad lines with bright $^{13}$CO lines and $^{12}$CO v=2--1 lines are
at the top and spectra dominated by narrow $^{12}$CO v=1--0 lines at the
bottom. \cite{Blake04} recently published a survey of CO
ro-vibrational emission lines from Herbig Ae stars, i.e., somewhat
older and more massive objects than those presented here. They find
that a significant part of the line emission must be excited by
resonance flourescence and that the line widths correlate well with
disk inclination. An interesting question is whether the emission
lines observed toward embedded objects exhibit the same
characteristics. The least complex spectra with narrow lines and weak $^{13}$CO lines and $^{12}$CO v=2--1 lines resemble the older Herbig Ae disks the most. However, some of the spectra of embedded
sources show $^{13}$CO lines and $^{12}$CO v=2--1 lines which are almost as bright at the
$^{12}$CO fundamental lines. This indicates both high optical depths as well as high excitation temperatures in excess of 700\,K. In addition, the de-reddened luminosity in the ro-vibrational 
CO band is in some cases 1-2 orders of magnitudes higher in embedded sources compared to the  Herbig Ae disks of \cite{Blake04}, although the intrinsic scatter is large. This suggests that a different
mechanism may produce the ro-vibrational emission in embedded stars. One possibility is cooling emission excited by accretion activity as proposed by \cite{Pontoppidan02}.

\section{Mapping ices on 1000\,AU scales in a protostellar envelope}

A new possibility in mid-infrared spectroscopy is the spatial mapping of ice abundances in
protostellar regions \cite{Pontoppidan04}. This has been facilitated mainly by the increase in sensitivity of both ground- and space-based infrared spectrometers rather than by developments in spectral resolution. However, high resolution spectroscopy in principle allows for a simultaneous determination
of the ice-to-gas ratio along a specific line of sight for the chemical species observed; the ice column density can be obtained from a stretching mode ice band, while the gas column density can be determined from absorption in the associated ro-vibrational band. This has been
taken advantage of in the past for isolated lines of sight, e.g.\ \cite{Boonman03}\cite{Thi02}. In order to unambiguously determine gas column densities from ro-vibrational absorption lines, it is necessary to spectrally resolve the lines. Since lines from cold molecular gas will have widths of less than 5\,$\rm km\,s^{-1}$, spectroscopy at $R=100\,000$ is in principle necessary. 
 
\begin{figure}[ht]
\begin{center}
\includegraphics[width=.8\textwidth]{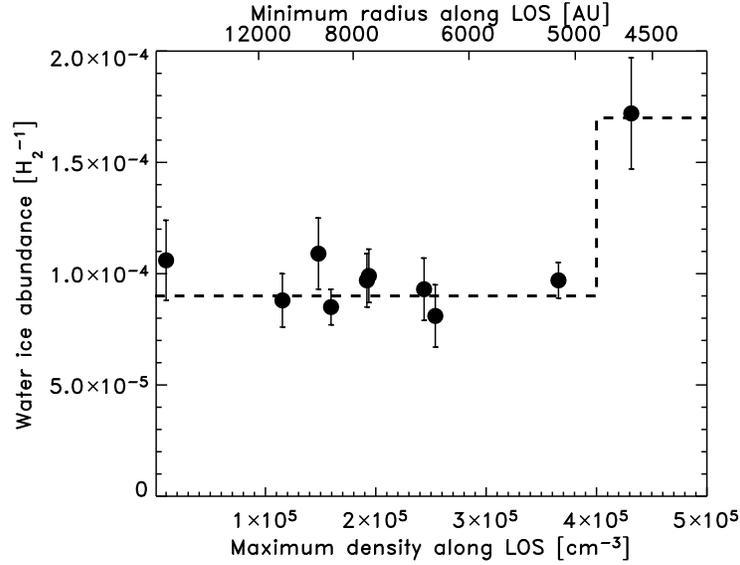}
\end{center}
\caption[]{The abundance of water ice with respect to H$_2$ in the envelope of the class 0 protostar SMM 4 in the Serpens cloud core as a function of distance to the envelope center. The dashed line indicates the location of an apparent jump in the abundance of water ice. }
\label{icemap}
\end{figure}

Fig. \ref{icemap} shows the water ice abundance as a function of distance to the center of the class 0 protostar SMM 4 in the Serpens cloud core. The abundances were determined using observations of the 3.01\,$\mu$m water ice band toward sources located in a small stellar cluster embedded in the envelope of SMM 4. The observed sources are spaced only 1000--2000\,AU apart, and the resulting water ice map thus has a spatial resolution two orders of magnitude better than any previous maps. The abundance of water ice remains constant throughout most of the envelope, but
the point closest to the center of SMM 4 shows an increase in abundance by a factor of 1.7. This is consistent with recent models predicting a jump in gas phase depletion at the density where the freeze-out timescale becomes smaller than the collapse timescale \cite{Joergensen04}. Similar observations of the 3.53\,$\mu$m methanol ice band shows that the abundance of methanol ice is enhanced by a factor of 10 to an abundance of $3\times 10^{-5}$ with respect to H$_2$ relative to the surrounding cloud medium in Serpens. The cause of this enhancement of the methanol ice abundance is presently not known.

\section{Acknowledgments}

The VLT-ISAAC survey was carried out in close collaboration with E.\
Dartois, W.F. Thi, L.\ d'Hendecourt, A.\ Boogert, H. Fraser,
S. Bisschop, W. Schutte and A. Tielens. Research in Astrochemistry in
Leiden is supported by grants from the National Research School for
Astronomy (NOVA) and a Spinoza grant from NWO.

%INDEX%%%%%%%%%%%%%%%%%%%%%%%%%%%%%%%%%%%%%%%%%%%%%%%%%%%%%%%%%%%%%%%
% Please check with the editor of your book whether he plans to
% include a "mutual" subject index - if so, please code your entries
% in the standard syntax. For your own purposes you may print your
% "personal" index by using the following commands:
%
%\clearpage
%\addcontentsline{toc}{section}{Index}
%\flushbottom
%\printindex
%%%%%%%%%%%%%%%%%%%%%%%%%%%%%%%%%%%%%%%%%%%%%%%%%%%%%%%%%%%%%%%%%%%%%

\end{document}